\documentclass[usenatbib,usegraphicx]{mn2e}
\usepackage{bethmacros}

\title[The Origin and Properties of Intracluster Stars in a Rich Cluster]{The Origin and Properties of Intracluster Stars in a Rich Cluster}
\author[B. ~Willman, F. ~Governato, J. ~Wadsley, T. ~Quinn]{Beth Willman$^{1}$\thanks{Email: willman@astro.washington.edu}, Fabio Governato$^{2,3}$\thanks{Brooks Fellow}, James Wadsley$^{4}$ and Thomas Quinn$^{2}$ \\
$^1$Center for Cosmology and Particle Physics, Department of Physics, New York University, 4 Washington Place, New York, NY 10003, USA  \\
$^2$Department of Astronomy, University of Washington, Box 351580, Seattle, WA 98195, USA \\
$^3$Osservatorio Astronomico Di Brera, via Brera 28, Milano, Italy\\
$^4$Department of Physics and Astronomy, McMaster University, Hamilton, Ontario L88 4M1, Canada}
\begin{document}
\maketitle


\begin{abstract}
We use a multi million particle N-body + SPH simulation to follow the
formation of a rich galaxy cluster in a $\Lambda$CDM cosmology, with
the goal of understanding the origin and properties of intracluster
stars. The simulation includes gas cooling, star formation, the
effects of a uniform ultraviolet background and feedback from
supernovae.  Halos that host galaxies as faint as M$_R$ = -19.0 are
resolved by this simulation, which includes 85\% of the total galaxy
luminosity in a rich cluster. We find that the accumulation of
intracluster light (ICL) is an ongoing process, linked to infall and
stripping events.  The unbound star fraction varies with time between
10\% and 22\% of the total amount of cluster stars, with an overall
trend to increase with time.  The fraction is 20$\%$ at z = 0,
consistent with observations of galaxy clusters. The surface
brightness profile of the cD shows an excess compared to a de
Vaucouleur profile near 200 kpc, which is also consistent with
observations.  Both massive and small galaxies contribute to the
formation of the ICL, with stars stripped preferentially from the
outer, lower metallicity, parts of their stellar distributions.
Simulated observations of planetary nebulae (PNe) show significant
substructure in velocity space, tracing separate streams of stripped
intracluster stars. Despite an unrelaxed distribution, individual
intracluster PNe might be useful mass tracers if more than 5 fields at
a range of radii have measured line-of-sight velocities, where an
accurate mass calculation depends more on the number of fields than
the number of PNe measured per field.  However, the orbits of IC stars
are more anisotropic than those of galaxies or dark matter, which
leads to a systematic underestimate of cluster mass relative to that
calculated with galaxies, if not accounted for in dynamical models.
Overall, the properties of ICL formed in a hierarchical scenario are
in good agreement with current observations, supporting a model where
ICL originates from the dynamical evolution of galaxies in dense
environments. ICL should thus be ubiquitous in galaxy clusters.

\end{abstract}




\section{Introduction}
\label{intro}

For many years, the presence of intracluster (IC) stars has been well
documented by observations of diffuse light between cluster galaxies
(e.g. \citealp{zwicky51, gonzalez00, feldmeier02, feldmeier04, zibetti04,
krick04}).  Recently, numerous new searches for IC stars have
uncovered individual stars between cluster galaxies (PNe:
\citealp{theuns97, arnaboldi02, arnaboldi03, feldmeier98, feldmeier03,
okamura02}; red giants: \citealp{durrell02}; SNeIa:
\citealp{galyam03}), and individual tidal streams in clusters
\citep{trentham98,gregg98,calcaneo00,feldmeier02}. These studies, and
others, all measure an amount of intracluster light that is between
10$\%$ and 50$\%$ of the total cluster luminosity, suggesting that
intracluster stars are a generic feature of galaxy clusters.

The observable properties of intracluster stars are likely closely
linked to the dynamical history of galaxy clusters, as IC stars are a
natural by-product of cluster evolution in hierarchical models such
as $\Lambda$CDM. Numerical simulations have demonstrated that the
dynamical evolution of galaxies in clusters is mainly dictated by
global gravitational tides and fast encounters between galaxies that
alter their morphologies and strip them of their dark matter
halos \citep{moore96,dubinski98,ghigna98,calcaneo00,mayer01}.  For
example, \citet{moore96} studied the dynamical evolution of individual
galaxies inside a realistic cluster potential to show that, depending
on their internal structure, they might undergo significant stripping
of their stellar material. As galaxies get stripped, stars in their
tidal streams carry memory of the parent galaxy's orbit for several
dynamical times.  Spatial streams seen in Hydra and Coma have thus
been used to trace past dynamical interactions \citep{calcaneo00}. 


As stripped stars become mixed into the global cluster potential, they
still bear signatures of the cluster's accretion and dynamical
history, although the extent of this is uncertain.  Thus far,
numerical studies that addressed such uncertainties have focused on
the z = 0 properties of intracluster stars.  Dubinski (1998) used an
approach where disk/halo galaxy models were substituted for dark
matter halos in a cluster simulation in a critical universe to show
that a significant fraction of stars end up unbound to individual
galaxies.  Napolitano et al. (2003) compared a collisionless N-body
simulation of the formation of a galaxy cluster with recent
observations of Virgo IC PNe. A simple scheme to identify a diffuse
stellar component in the simulation showed that intracluster stars are
largely unrelaxed in velocity and are clustered on 50 kpc scales at
500 kpc from the cluster center, in good agreement with the observed
clustering properties of the diffuse population of PNe in Virgo.
Recently, \citet{sommerlarsen04} looked at the z = 0 photometric
properties and metallicity distribution of IC stars in a cosmological
simulation, and found them in good agreement with observations.
\citet{murante04} studied clusters in a large cosmological simulation
and found fractions of unbound cluster stars ranging from 10 - 50$\%$.
However, both \citet{sommerlarsen04} and \citet{murante04}'s spatial
resolution was much coarser than the average scale size of cluster
galaxies, which should overproduce intracluster stars, due to overly
efficient stripping.

The numerous remaining uncertainties in the expected properties of IC
stars leave much room for improvement in this burgeoning field of
study.  For example, no consensus exists on when and how many IC stars
should actually be produced in a case where the entire
population of a forming galaxy cluster is realistically simulated. Nor
is there consensus on whether IC stars originate mostly from the
stripped outer parts of bright galaxies or rather from a population of
tidally destroyed low surface brightness galaxies.  It is also unclear
whether observed cluster to cluster variations in measured IC
fractions and radial profiles are due to different cluster formation
histories and dynamical states, rather than a cosmic scatter in the
expected properties of IC stars.

Furthermore, little has been done theoretically to predict the
dynamics of the intracluster population; there is only one published
analysis of the velocity distribution of IC PNe
\citep{sommerlarsen04}. Positions and velocities have been measured
for hundreds of IC planetary nebulae in Virgo
(e.g. \citealt{feldmeier03,arnaboldi02}) making it possible to use PNe
to measure cluster mass, as ICPNe outnumber cluster galaxies
\citep{ford02}.  However, it is unclear if an accurate mass
measurement with PNe is feasible, due to the substructure found in
observations and simulations of IC stellar populations.


 In this paper, we address the outstanding questions above by using a
 simulation of a rich galaxy cluster in a realistic cosmological
 environment by using an unsurpassed coupling of high spatial and mass
 resolution.  This allows us to follow the star formation histories of
 individual galaxies as faint as L$_{*}$/10 for the first time, a
 crucial requirement to resolve the formation of the large majority of
 the stars formed in the cluster. Nearly $90\%$ of galaxy light in the
 Coma cluster resides in galaxies with L$>$ L$_{*}/10$
 \citep{mobasher03}. Also, due to a spatial resolution higher than
 most previous analyses, our simulations produced a realistic
 distribution of galaxy sizes as a function of stellar mass and dark
 matter halo mass.  This requirement is fundamental for a faithful
 description of the stripping processes in the dense cluster
 environment.  Also, although simulations with a more ad hoc approach
 to tracing stars produce worthwhile results, a simulation that
 includes star formation produces a more accurate model of ICL as a
 continuous, ongoing process.
 
The aims of this paper are: a) to shed light on the uncertain origin
and properties of intracluster stars in a $\Lambda$CDM Universe and how they
relate to the dynamical state of the cluster and b) test their use as
dynamical tracers of a galaxy cluster potential.  In \S2, we describe
the simulation; in \S3, we describe the origin and evolution of the IC
population; in \S4, we discuss IC stars as tracers of the cluster's
dynamical history; and in \S5, we evaluate the utility of IC PNe as
tracers of cluster mass.

\section{The Simulation}
\label{sec:simulation}
\subsection{GASOLINE}
We selected a candidate Coma cluster from an existing low resolution,
dark matter only simulation in a concordance ($\Lambda$=0.7,
$\Omega_0$=0.3, $\sigma_8$=1) cosmology and resimulated it at higher
resolution using the volume renormalization technique \citep{katz93}.
To perform this simulation, we used GASOLINE \citep{wadsley03}, a
smooth particle hydrodynamic (SPH), parallel treecode that is both
spatially and temporally adaptive.  The version of GASOLINE we used
implemented i) Compton and radiative cooling, ii) star formation and
'minimal' supernova feedback as described by \citet{katz96}, and iii)
a UV background following \citet{haardt96}.  The opening angle,
$\theta$, was 0.5 until z = 2 and 0.7 thereafter, and the
time-stepping criterion, $\eta$, was 0.2, as in \citet{diemand04}.

We ran four different simulations of the same cluster:

- C1 is a dark matter only run that we use to evaluate our halo
completeness limit.  This is a crucial measure as stars can form
in poorly resolved halos but should not be included into our analysis.

- C2 is our fiducial run that includes gas cooling processes, star
  formation and SN feedback. The high resolution region of this
  simulation contains dark and star particles of $1.5 \times 10^9$ and
  $6.7 \times 10^7$ $M_{\odot}$ respectively, with a force softening
  of 3.75 kpc.  There are $2 \times 10^6$ particles within the
  cluster's virial radius at z $=$ 0.  We will discuss results from
  C2, unless noted otherwise.

- C2low is a lower resolution version of C2 with 8 times fewer particles.

- C3 is of equal resolution to C2, but includes the effects of
additional feedback introduced by setting the gas at an entropy larger
than 5 keV cm$^{2}$ at z = 3 to about 50 keV cm$^{2}$, as measured by
X-ray observations of clusters and groups (Borgani \etal 2002).

All four clusters have R$_{200}$, the radius enclosing an average
$\delta \rho$/$\rho$ = 200, of 2.2 Mpc, M$_{200} \sim 1.2 \cdot
10^{15}$ M$_{\odot}$, and a 1D velocity dispersion of 1000 km
sec$^{-1}$.  These are similar to the observed properties of the Coma
cluster \citep{hughes89}.

Table~\ref{tab:simulations} gives the number of particles, particle
masses, and force softenings of each simulation.  The stellar masses
given in the table are the maximum mass of a star within each z = 0
cluster.

\begin{table}
\begin{tabular}{cccccc}
\hline
run & N$_{dark}$ & N$_{star}$ & M$_{dark}$ & M$_{star}$ & $\epsilon$ \\
 & \multicolumn{2}{c}{within $R_{vir}$} & \multicolumn{2}{c}{M$_{\odot}$ per particle} & kpc \\ \hline
C1 & 680814 & -- & 1.7 $\cdot$ 10$^9$  & -- &  3.75 \\ \hline
C2 & 688754 & 868995 & 1.5 $\cdot$ 10$^9$ & 7.2 $\cdot$ 10$^7$  & 3.75 \\ \hline
C2low & 85570 & 143326 &  1.2 $\cdot$ 10$^{10}$ & 8.3 $\cdot$ 10$^8$ & 7.5 \\ \hline
C3 & 679574 & 516261 & 1.5 $\cdot$ 10$^9$ & 7.2 $\cdot$ 10$^7$ & 3.75 \\ \hline
\label{tab:simulations}
\end{tabular}
\caption{Simulation Parameters}
\end{table}

\subsection{Simulated Cluster Properties}

\begin{figure}
\begin{center}
\resizebox{8cm}{!}{\includegraphics{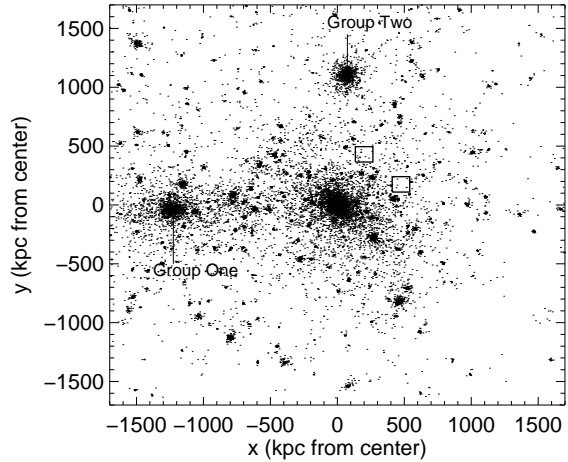}}\\
\end{center}
\caption{The spatial distribution of stars in the simulated Coma-like
galaxy cluster, C2.  Only 2$\%$ of the star particles are displayed,
for clarity.  The outlined squares show the size of a 0.5 x 0.5
deg$^2$ field for the cluster at a distance of 15 Mpc, the distance of
Virgo.}
\label{fig:clusterimage}
\end{figure}

Figure~\ref{fig:clusterimage} shows the spatial distribution of the
cluster stars at z = 0, represented by a random selection of 2$\%$ of
all stars.  The clear asymmetry in this distribution suggests that the
galaxy population is unrelaxed, which we show in \S3 to be the case,
due to several major accretion events at low redshift. The outlined
black squares show the size of a 0.5 x 0.5 deg$^2$ field at 15 Mpc,
the distance of Virgo.  We discuss the properties of stars in these
fiducial fields in \S4.  There are two large groups in the cluster:
Group One is a Virgo sized group that entered the cluster 1 -- 2 Gyr
ago.  Group Two is the size of Fornax, and entered the cluster 5 Gyr
ago.


20\% of the simulated cluster's stars are unbound to individual
galaxies at z = 0, consistent with the values of 10 -- 50\% suggested
by observations (see \S3.3 for additional discussion).  To distinguish
stars gravitationally bound to individual galaxies from free-floating
stars, we used the
SKID\footnote{http://www-hpcc.astro.washington.edu/tools/skid.html}
halo finding algorithm on the gas and dark matter. We initially ran
SKID with a large linking length of 18 kpc, to minimize identifying
loosely bound stars at z $=$ 0 as unbound.  Because that choice of
linking length misses a few halos within the cD envelope, we also ran
SKID with a smaller linking length of 9 kpc and included the
additional galaxies in our sample.  Based on a visual evaluation of
the data, this combination produced the most robust identification of
halos at z = 0.  However, we found that our choice of linking length
does not substantially affect the results.  This procedure identified
650 galaxies within R$_{200}$.

Due to the weak implementation of feedback in the simulations, star
formation in dense gas regions proceeds unimpeded, resulting in a
stellar baryon fraction of 36$\%$, several times higher than the 6 --
12$\%$ suggested by observations of galaxy clusters
\citep{bell03b,balogh01,cole01}.  This ``cooling crisis'' is a well
known weakness of the current generation of cosmological SPH
simulations (Cole et al. 2001, Bell et al. 2003, Balogh et al. 2001,
Kay 2001).  Producing the correct amount of stars in a cosmological
simulation is a problem that depends critically on numerical
resolution and on the poorly known details of star formation and
feedback from SNe (Thacker \& Couchman 2000, Borgani \etal 2001),
although see \citet{springel03} and \citet{sommerlarsen04} for recent
improvements.

However, this overproduction of stars does not adversely affect our
results because star particles in C2 galaxies have a realistic
distribution, and are thus accurate tracers of the dynamically
stripped stellar population. If the simulated stellar halos are too
compact or too diffuse, then the stripping efficiency by dynamical
processes in the cluster would be under or overestimated,
respectively.  To determine the stellar distributions of galaxies in
C2, we fit de Vaucouleurs effective radii ($R_e$) to the 389 cluster
galaxies with stellar mass $> M_{limit}$ $(10^{9.9}
M_{\odot})$. $M_{limit}$ is the typical simulated stellar mass
contained in a halo just above the conservative resolution limit of
the simulation.  We determine the halo resolution limit by a
comparison with C1, the dark matter only run.  The circular velocity
function of C1 started to flatten $\sim v_c = 100$ km sec$^{-1}$,
although C2's did not flatten until 70 km sec$^{-1}$.  Halos less
massive than 100 km sec$^{-1}$ would suffer less efficient cooling and
may be incomplete due to numerical effects.  We thus selected v$_c$ =
100 km sec$^{-1}$ as a conservative resolution limit to be confident
of completeness in our analysis.  Nearly all galaxies with stellar
masses greater than $10^{9.9} M_{\odot}$ are hosted by halos with v$_c
>$ 100 km sec$^{-1}$.  Halos more massive than this limit contain
98\% of the simulated stellar mass.  We cut on stellar mass, rather
than velocity or dark matter mass, to facilitate matching each
simulated halo to a galaxy drawn from a Coma cluster luminosity
function (\S2.2.1).  

\begin{figure}
\begin{center}
\resizebox{8cm}{!}{\includegraphics{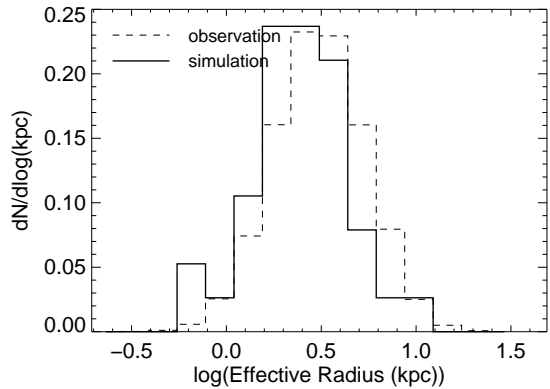}}\\
\end{center}
\caption{Histogram comparison of the distribution of de Vaucouleurs
effective radii for our simulated galaxies vs the distribution
measured by Bernardi et al. (2003) for 9000 early type galaxies in the SDSS.}
\label{fig:rehist}
\end{figure}

Figure~\ref{fig:rehist} shows the resulting distribution of the
simulated galaxies' effective radii, overplotted with the distribution
measured by \citet{bernardi03}, for 9000 early type galaxies in the
SDSS.  The $R_e$ distribution of C2 galaxies is indeed in the same
range and peaks near the same scale size as that observed by
\citet{bernardi03}. The simulated galaxies are systematically slightly
more compact than those observed, which may result in a lower unbound
star fraction than reality (see \S3.3).  However, our measured
distribution of scale sizes is a substantial improvement over existing
studies of IC stars.  We thus conclude that our simulated galaxies
have sufficiently realistic stellar profiles to be used to study the
production of intracluster stars in a rich cluster.

\subsubsection{The Galaxy Luminosity Function}
\begin{figure}
\begin{center}
\resizebox{8cm}{!}{\includegraphics{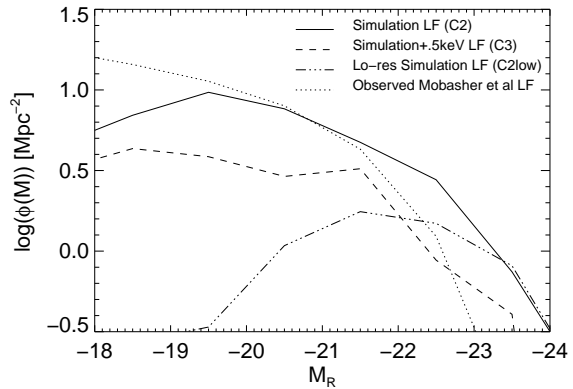}}\\
\end{center}
\caption{Original luminosity functions of simulated cluster galaxies
in three different runs: low resolution, high resolution, and high
resolution with an entropy injection of 0.5 keV at z = 3.  For
reference, the observed luminosity function of Mobasher et al (2003)
is overplotted.}
\label{fig:LForig}
\end{figure}

One of our main goals is to study what fraction of ICL originates from
galaxies of different masses.  Our simulation resolved a range of
4 decades in stellar mass; to determine how far down the galaxy
luminosity function (LF) this stellar mass range includes and to
evaluate which of our simulations to focus this study on, we compare
the luminosity functions of our simulations with that observed for the
Coma cluster. Figure~\ref{fig:LForig} shows a comparison between the
original LFs of C2, C2low, C3, and the luminosity function for the
entire Coma cluster as observed by \citet{mobasher03}.  To convert the
total stellar mass of each simulated cluster galaxy to an R magnitude,
we combine the \citet{secker97} observed R vs. B-R relationship for
galaxies in Coma with the \citet{bell01} (M/L)$_R$ vs (B-R) relation
with a scaled Salpeter IMF and \citet{cole00} galaxy evolution model.


As star particles formed in the simulation act as dynamical tracers,
it is necessary that our simulation has a galaxy LF as close as
possible to that observed for cluster galaxies.
Figure~\ref{fig:LForig} shows that C2 galaxies approximately follow a
Schecter function, although the brightest galaxies overproduce stars.
However, the lower resolution run, C2low, neither produces the correct
number of stars in massive halos nor resolves dark matter halos
smaller than those that host M$_R^{\star}$ galaxies. A run that
resolves such a small fraction of the luminosity function would
provide a very unrealistic and incomplete description of intracluster
star accumulation.  Although C3, the run with 0.5 keV particle$^{-1}$
energy injection, produces a reasonable luminosity function above
M$_R^{\star}$, the formation of galaxies fainter than M$_R^{\star}$ is
unrealistically suppressed.  This feedback recipe, while useful to
study the properties of the IGM \citep{borgani02}, is too crude to
accurately simulate galaxy formation within the cluster.  Based on
these comparisons, we choose to focus our analysis only on C2.

\begin{figure}
\begin{center}
\resizebox{8cm}{!}{\includegraphics{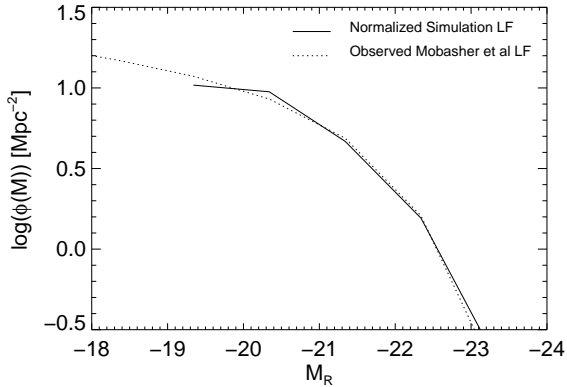}}\\
\end{center}
\caption{The luminosity function of simulated cluster galaxies after
normalization, as compared to the observed luminosity function of
Mobasher et al (2003).}
\label{fig:LFnorm}
\end{figure}

To make our results directly comparable with observations, we derive
corrected stellar masses and R magnitudes for each cluster galaxy by
normalizing to the observed \citet{mobasher03} luminosity function
(LF).  Although the mass-luminosity relation of C2 is not correct, the
basic correlation is correct: more stars form in more massive halos.
Since the correlation between stellar and halo mass is monotonic, and
we have a large number of halos, we are justified in using the
observed luminosity function of the Coma cluster to calibrate the
results of our star formation algorithm.  We perform the normalization
based on the 389 galaxies above the halo resolution limit (see above).
The integrated \citet{mobasher03} LF of the entire Coma cluster
contains 389 galaxies brighter than R = -19.0, which includes 86$\%$
of the total light they observed in cluster galaxies.  We thus create
rank ordered populations of 389 galaxies drawn from a Mobasher LF
cutoff at R = -19.0, which is 2.8 magnitudes fainter than
M$_R^{\star}$ of Coma galaxies.  We convert their M$_R$s to estimated
stellar masses by combining \citet{secker97}'s observed
color-magnitude relation for Coma with the \citet{bell01} color-(M/L)
relation used above. We then fit a function relating these stellar
masses to the rank ordered simulation stellar masses.  We convert
M$_R$ to stellar mass, rather than just matching simulated stellar
mass to M$_R$, so that we can normalize the mass of each individual
star in the simulation.  Figure \ref{fig:LFnorm} shows the normalized
luminosity function of our cluster.  After normalizing, 10$\%$ of
cluster baryon mass is in stars, vs. 36$\%$ pre-normalization.

We used the stellar mass normalization derived at z = 0 for cluster
galaxies at all redshifts up to z = 1.1, correcting the derived
luminosities for passive evolution.

\section{Origin and Evolution of the Intracluster Stars}
\label{sec:origin}

Because this cosmological simulation includes a large dynamic range of
galaxies, we can derive the contribution of galaxies to the IC
population, as a function of their absolute magnitude.  In this
section, we determine the origin of IC stars and use our result to
predict and interpret observed colors and metallicities of IC stars.
We also determine the evolution of the intracluster star fraction and
evaluate the extent to which it is a function of the cluster's
dynamical state.

\subsection{Determining the Progenitor Halos of the Intracluster Stars}
We traced all of the stars unbound at z = 0 back until the epoch when
they were last bound to a galaxy. As 85$\%$ of unbound stars were
stripped since z = 1.1, we chose to only trace to that epoch.  To
reduce errors in assigning stars to their progenitor halos, we $(i)$
do not allow stars to be ``stripped'' from the cD galaxy, but rather
traced them back to their original halos and $(ii)$ require that a
star is bound to its progenitor halo for at least two consecutive
simulation outputs.  We normalize the mass of each stripped star by
the same factor that its progenitor's mass was normalized, as
determined by the technique described in \S2.2.1.  We use these
normalized masses in the subsequent results presented in this paper.

\subsection{Determining the z = 0 Luminosities of the Progenitor Halos}
Thus far, observational studies of IC starlight have been limited to
relatively nearby, low redshift, clusters.  Therefore, we determine
the z = 0 luminosity of the galaxy each progenitor evolves into.  We
trace the member stars of each progenitor galaxy forward in time to
determine the halos they reside in at z = 0.  We define the z = 0
galaxy, G$_f$, of each progenitor, G$_i$, as the z = 0 galaxy that
contains the highest fraction of G$_i$'s stars.  If the majority of
G$_i$'s stars are unbound at z = 0, then G$_f$ must contain $\geq
40\%$ of G$_i$'s 10 most bound stars.  If not, the progenitor halo is
considered to be destroyed by z = 0.  Only 1\% of unbound stars from
galaxies brighter than M$_R$ = -19, were stripped from galaxies that
were ultimately completely disrupted, and 9\% of unbound stars were
stripped from galaxies that ultimately merged with the cD.

\subsection{Evolution}
Observations of diffuse intracluster light have suggested that the
majority of the IC star population in some clusters was accumulated
early in the formation of the cluster \citep{gonzalez00}.  Authors
have also suggested that the fraction of unbound stars in galaxy
clusters may be linked to the dynamical state of the cluster
\citep{feldmeier02}.  In this section, we compare these past results
to the evolution and fraction of intracluster stars in C2.

\begin{figure}
\begin{center}
\resizebox{8cm}{!}{\includegraphics{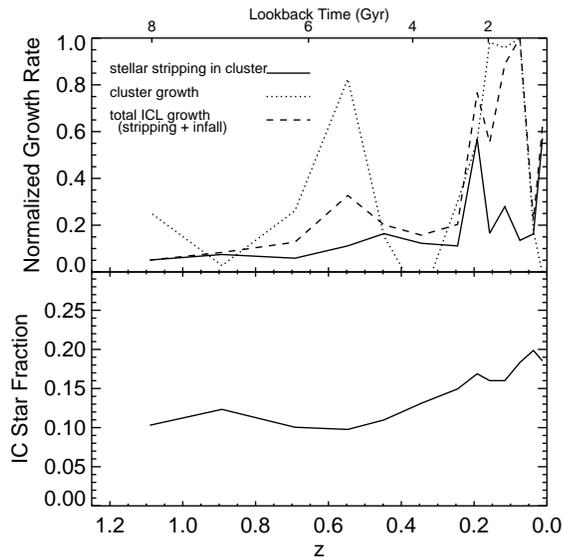}}\\
\end{center}
\caption{Top panel: Time evolution of the stripping and cluster growth
rates compared with the time evolution of the fraction of cluster
stars in the ICM.  The values of stellar stripping and total ICL
growth are both normalized to the maximum ICL growth rate.  The values
of cluster growth rate are normalized to the maximum cluster growth
rate.  Bottom panel: The time evolution of the unbound star fraction.}
\label{fig:evolution}
\end{figure}

Figure \ref{fig:evolution} shows the evolution of C2's intracluster
population back to z = 1.1.  The top panel compares the growth rate of
the intracluster stars with the overall cluster growth rate.  The
peaks at z = 0.55 and z = 0.2 to 0.1 correspond to the infall of large
groups, ranging in size from Fornax to Virgo. The growth of the IC
population and the cluster as a whole are clearly linked and primarily
occur by an ongoing series of events, rather than by a steady
process. Stars are added to the intracluster halo both via stripping
processes within the cluster environment and via the infall of large
galaxy groups that already contain unbound stars.

The bottom panel of Figure \ref{fig:evolution} shows the fraction of
cluster stars that are unbound as a function of time. The IC star
fraction in C2 displays a trend to increase with time.  This trend is
due to the continuous stripping of stars from cluster galaxies as they
orbit.  Although the infall of large galaxy groups adds stars to the
IC population, the actual IC fraction is not necessarily linked to the
dynamical state of the cluster. For example, the cluster underwent a
substantial accretion event at z = .55, but did not have an enhanced
IC fraction as a result.  This event increased the mass of the
cluster's IC population primarily via stars already unbound within the
infalling group, rather than by stripping within the cluster.  The
ratio of added IC star mass to added total cluster mass was thus not
larger than the ratio already present in the cluster, and thus did not
increase the unbound star fraction.

Since z = 1.1, the IC fraction varied between a minimum of 10$\%$ and
a maximum of 22$\%$, which closely matches the range in IC star
fractions estimated for Virgo and several Abell clusters
\citep{feldmeier98,feldmeier04}, although is less than the fraction
estimated for Coma \citep{bernstein95} and a rich Abell cluster
\citep{feldmeier02}.  This difference could be due to the facts that
(1) the simulated galaxies are a bit more compact than those observed,
and (2) the fraction we measure represents a lower limit to the
fraction of cluster stars that are unbound, because can not include
stars stripped from galaxies fainter than -19 in our analysis.

Although stars are continuously added to the intracluster population,
the simulated cluster accumulated 50$\%$ of its IC stars within the
last 2 Gyr. The upper panel of Figure \ref{fig:evolution} shows that
this recent increase in the number of IC stars was connected to a
substantial accretion event experienced by the cluster between 1 and 2
Gyr ago.  The mean dynamical time of this cluster is $(3 \pi / 16
G\rho)^{-1/2} \sim$ 2 Gyr, thus the cluster is not in a dynamically
relaxed state, as suggested by the asymmetric spatial distribution of
the cluster seen in Figure 1.  The variation of IC fraction with time
in just one cluster suggests that there is no universal IC
fraction.  

\subsection{Origin}
\begin{figure}
\begin{center}
\resizebox{8cm}{!}{\includegraphics{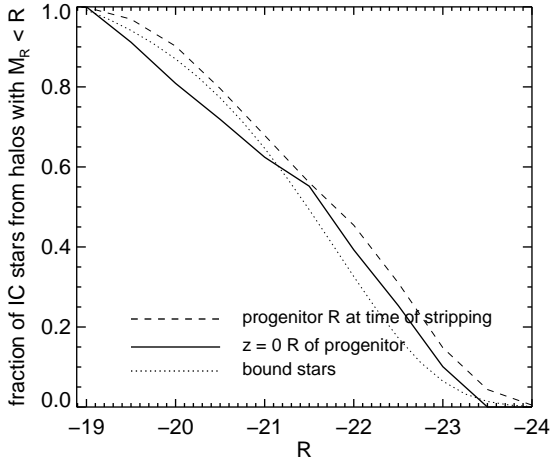}}\\
\end{center}
\caption{The fraction of intracluster stars stripped from halos as a
function of their progenitor's absolute magnitude, both at the time of
stripping and traced forward to z = 0.  The fraction of stars bound to
halos with M$_R < R$ is also overplotted.}
\label{fig:origin}
\end{figure}

Figure \ref{fig:origin} shows the fraction of IC stars stripped as a
function of the absolute magnitude of their progenitor halo, both at
the time of stripping, z$_{strip}$, and at z = 0.  9\% of stars are
stripped from galaxies that become fainter than -19 by z = 0. The
cumulative fraction of stars bound to halos with M$_R$ $<$ R at z = 0
is also overplotted.  The fact that the fraction of stars from halos
with M$_R$ $<$ R at the time of stripping closely follows the fraction
of stars contained in halos (plus an offset in M$_R$) shows that
stripping efficiency is roughly constant with stellar mass.  The
offset in M$_R$ results from our accounting for passive evolution when
determining M$_R$ at z $>$ 0.  The z = 0 progenitor M$_R$ distribution
has a different shape than the z = z$_{strip}$ distribution because
stripped galaxies that are less massive than M$^*$ often merge into
more massive galaxies.

We find that massive galaxies contribute substantially to the IC
population.  Figure \ref{fig:origin} also shows that 50$\%$ of
free-floating cluster stars were stripped from progenitor galaxies
with luminosities of M$_R^{\star}$ or brighter.  Because stars are
preferentially stripped from the outer regions of galaxies, IC stars
should bear the properties of stars in the outskirts of galaxies from
which they were stripped.  Assuming passive evolution, these should be
similar to the z = 0 properties of stars in the outskirts of
progenitor galaxies.  

Stars in the outskirts of L$^*$ and brighter galaxies have photometric
properties that differ from those of the stars producing the majority
of galaxy light, due to negative metallicity gradients
\citep{peletier90}.  Therefore, stars stripped from the outer regions
of the most massive galaxies could have the same colors and
metallicities of an 'average' star in an intermediate luminosity
galaxy.  For example, the outer regions of elliptical galaxies have
been observed to have (B-R) that are $\sim$ 0.1 - 0.15 magnitudes
bluer than the inner regions \citep{peletier90}.  \citet{gladders98}'s
observations of the red sequences of galaxy clusters showed that the
average (B-R) of dwarf galaxies are $\sim$ 0.3 magnitudes bluer than
the average (B-R) of the most luminous cluster galaxies.  Therefore,
we expect many IC stars to have colors in common with intermediate
luminosity galaxies (redder than the dwarfs, bluer than the most
luminous), although they have been stripped from galaxies brighter
than M$^*$.  This prediction is consistent with the \citet{durrell02}
measurement of -0.8 $<$ [Fe/H] $<$ -0.2 for the range of metallicities
of intracluster stars in Virgo.  This metallicity range agrees with
the mean metallicities of intermediate luminosity galaxies.

Low surface brightness (LSB) and dwarf galaxies fall below the
resolution limit of our simulation. We thus likely underestimate the
fraction of unbound stars.  Could LSBs and dwarfs actually be the
population contributing most of the IC stars?  \citet{durrell02}
presented the example, that although dwarf galaxies contribute only
$\sim 10\%$ of the light of Virgo, the IC stars contain $\sim 20\%$ of
the light in Virgo.  This means that for the IC stars to have all come
from dwarfs, the number of dwarfs must have once been 3 times their
present number. This extreme scenario is not required, as our results
show that massive galaxies produce substantial ICL. Simulations that
include galaxies with a more realistic distribution of profiles and/or
include galaxies even further down the luminosity function are
necessary to know the stripping efficiency as a function of luminosity
more precisely than we present here.

\section{Intracluster Stars as Tracers of Dynamical History}

In \S3, we showed that the cluster is dynamically unrelaxed and has
accumulated a substantial fraction of its IC population over the last
few Gyrs.  In this section, we investigate the extent to which this
dynamical history is reflected in the observable spatial and velocity
distributions of the cluster's stars.

\subsection{Spatial Distribution}
\label{sec:amount}
 
\citet{gregg98} and \citet{calcaneo00} identified cold stellar streams
in Coma with $\mu_R \sim 26.0$.  Such streams are evidence for recent
stripping events, because stars have not had sufficient time to get
mixed into the global cluster potential.  C2 underwent several major
stripping events within the last 2 Gyr, associated with the infall of
a Virgo sized group ('Group One' in Figure 1).  As a result, the
cluster exhibits a wide stream of unbound stars that extends for
hundreds of kpc between the cD galaxy and Group One, along the line
that it recently crossed the cD.  The average surface brightness of
the unbound stars in this large feature is $\sim \mu_R = 26.5$.  We
find no other streams in C2, but we likely underestimate the
prevalence of streams that result from such an accretion event,
because the galaxies in this simulation are all spheroidal.
Therefore, they all produce streams with internal velocity dispersions
($\sigma_{stream}$) that are comparable to v$_{rot}$, rather than the
much colder streams produced by disk galaxies ($\sigma_{stream} \sim
\sigma_{disk}$).

\begin{figure}
\begin{center}
\resizebox{8cm}{!}{\includegraphics{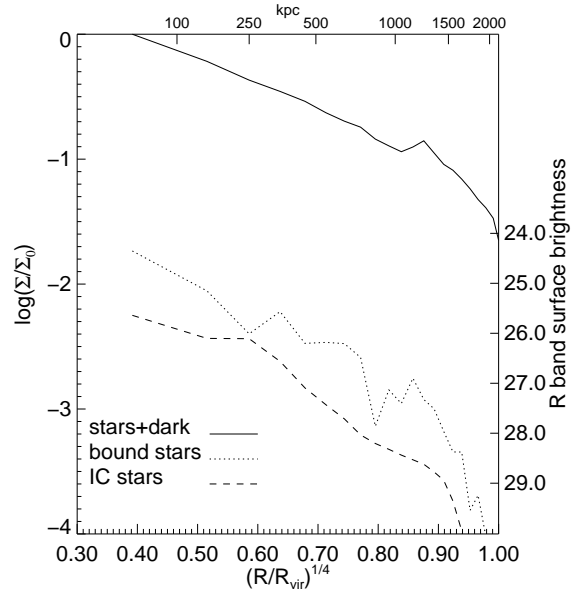}}\\
\end{center}
\caption{Comparison of the surface density profiles of: the stars
bound to galaxies, the intracluster stars, and all stars +
dark matter.  These projected profiles only include stars within the
virial radius of the galaxy cluster.}
\label{fig:profiles}
\end{figure}

The surface brightness profile of C2's IC stars is in agreement with
that observed in previous studies of both Coma and of other Abell
clusters.  Figure \ref{fig:profiles} displays the azimuthally averaged
surface density profiles of: the galaxy cluster (all stars+dark
matter), the stars bound to individual galaxies, and the intracluster
stars.  The profiles have been normalized to the average surface
density of the cluster within the central 100 kpc.  Although the
recent merger event results in an asymmetric IC star distribution,
their azimuthally averaged surface density profile beyond 250 kpc
follows that of the galaxies, as also seen in two of the Abell
clusters observed by \citet{feldmeier04}. Within 250 kpc, IC
stars have a flatter profile than bound stars (including cD stars), at
difference with the results of \citet{murante04}.  In fact, our
simulated IC population has a nearly flat surface brightness profiles
between 50 and 250 kpc.  Beyond 50 kpc, most of the light in the bound
stellar profile is not from cD stars. The average R-band surface
brightness of intracluster stars 250 kpc from C2's center is 26.2,
which agrees with that measured for Coma by \citet{bernstein95}.

\begin{figure}
\begin{center}
\resizebox{8cm}{!}{\includegraphics{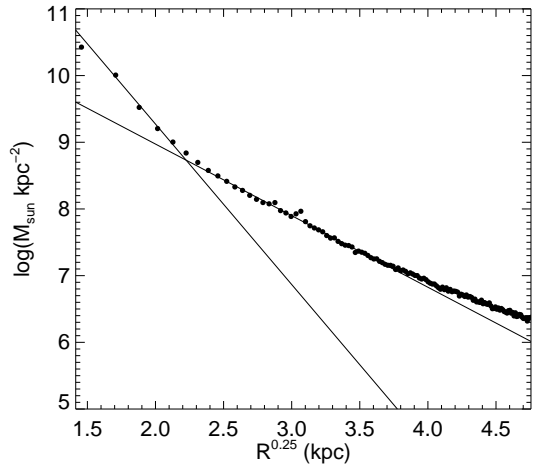}}\\
\end{center}
\caption{Surface density profile of the cD galaxy. Two R$^{1/4}$ fits
are overplotted; one was fit from 5 -- 20 kpc and one was fit from 40
-- 200 kpc.}
\label{fig:cDprofile}
\end{figure}

Figure~\ref{fig:cDprofile} shows the azimuthally averaged total
projected stellar profile of the cD galaxy with 2 de Vaucouleurs fits
overplotted.  The profile is inconsistent with a typical cD profile of
a single R$^{1/4}$ law, either with or without an excess at large
radii.  The unusually steep slope in the inner 25 kpc is likely due to
the unrealistic cooling flow and ongoing star formation at the center
of our cD, and is not seen in the profiles of the other simulation
galaxies. Although the radius ranges fit are arbitrary, the fits
highlight a true flattening in the outer envelope of the cD galaxy,
around 200 kpc.  This flattening can not directly be due to Group One
or Two, as they are a Mpc from the cluster center.  This excess above
a $R^{1/4}$ profile is similar to that seen by \citet{feldmeier02} for
the outer profile ($>$ 80 kpc) of the cD galaxy in the rich galaxy
cluster A1413.  There is building evidence that such envelopes may
even be ubiquitous features of cD galaxies of rich clusters (Gonzalez
et al., in preparation). The fact that C2's cD displays an excess over
R$^{1/4}$ at large R supports \citet{feldmeier02}'s suggestion that
such envelopes are a signature of recent cluster growth.  An
intracluster population formed primarily during the initial cluster
collapse differs from that of an intracluster population accumulated
slowly over time via tidal processes, and is expected to follow an
R$^{1/4}$ out to even larger radii than seen around C2
\citep{merritt84,dubinski98}.


\subsection{Substructure in Velocity Space}

We looked at the phase-space distribution of ``planetary nebulae'' (PNe)
in four 120 x 120 kpc fields (30$'$ x 30$'$ at a distance of 15 Mpc),
selected by eye to lie outside the cluster galaxies.  We used the
luminosity-specific Planetary Nebula number density of $\alpha_{1.0}$
= 9.4 $\times$ 10$^{-9}$ PNe L$_{\odot}^{-1}$ \citep{ciardullo89} to
randomly select a number of stars in each field representative
of the number of Planetary Nebulae that existing spectroscopic surveys
would observe.

\begin{figure*}
\begin{center}
\resizebox{8cm}{!}{\includegraphics{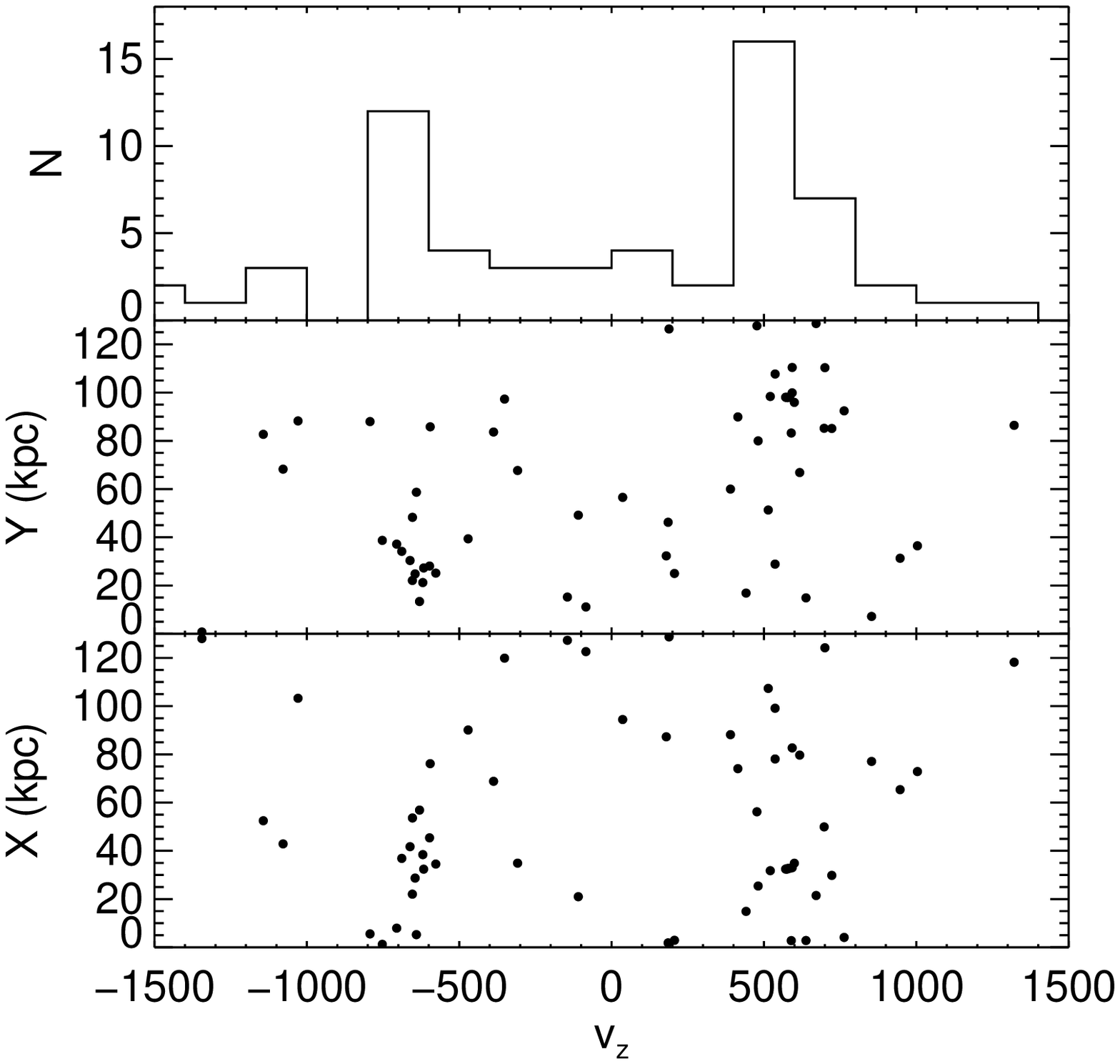}}\resizebox{8cm}{!}{\includegraphics{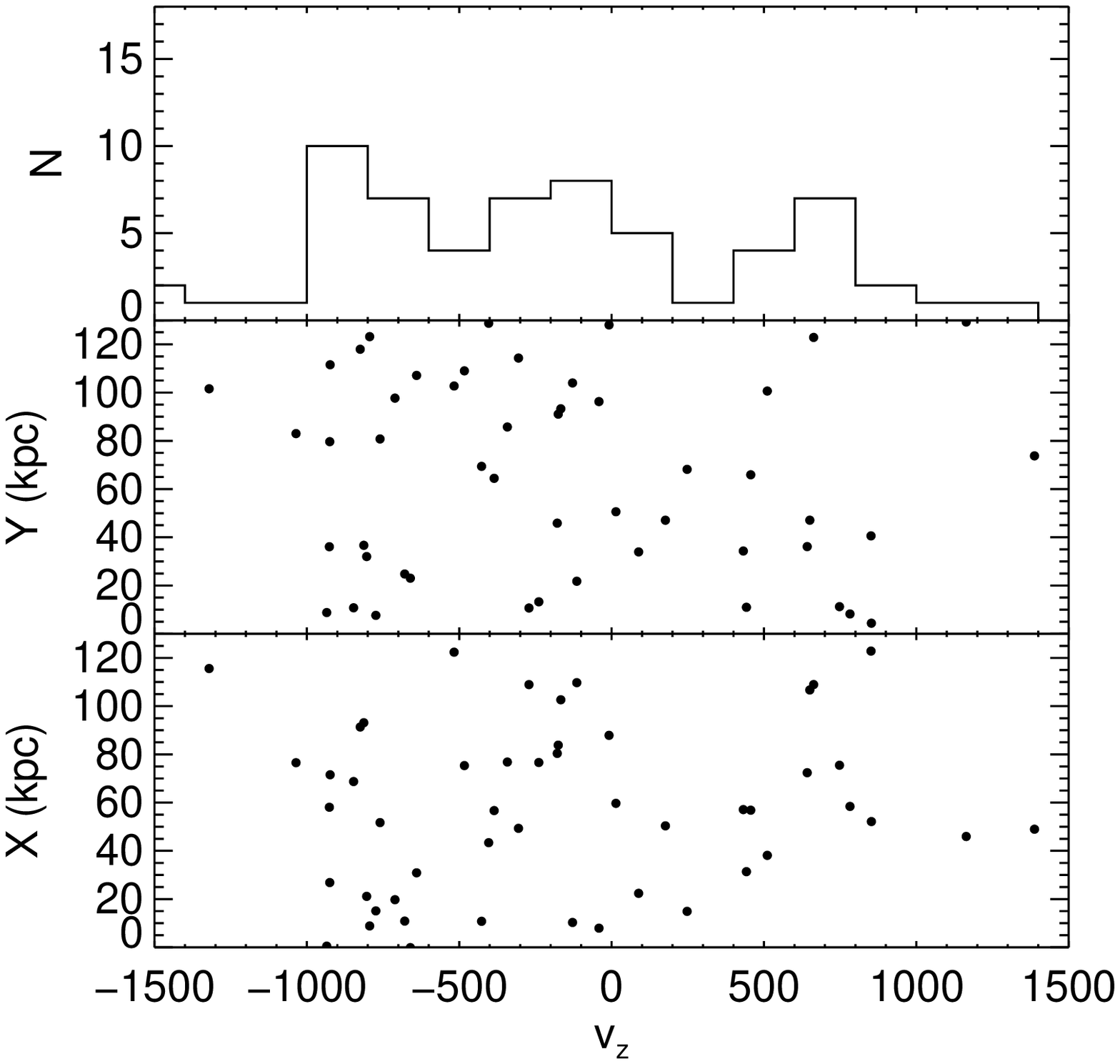}}\\
\end{center}
\caption{The spatial and velocity distribution of stars in the 2
'empty' fields in Figure~\ref{fig:clusterimage}, located 500 kpc from
the cluster center .}
\label{fig:velfield}
\end{figure*}

Figure~\ref{fig:velfield} shows the observable phase space
distributions and line-of-sight velocity histograms of PNe in the two
of the four fields plotted on Figure~\ref{fig:clusterimage}. The fact
that these 'observed' fields display both visible phase space
correlations and non-Gaussian distributions of line of sight
velocities reflects the young dynamical age of the IC population, also
found by \citet{napolitano03} when they applied the same analysis to
their simulations. The small scale phase-space substructure seen in
our cluster's stars represents a lower limit on the actual
substructure one would expect to observe, because all of the galaxies
in our cluster are elliptical, so stellar streams have higher internal
velocity dispersions, on average, and are thus more easily mixed than
those that would result from a population that contains disk galaxies.

\subsection{The Anisotropy Profile of the IC Stars}

\begin{figure}
\begin{center}
\resizebox{8cm}{!}{\includegraphics{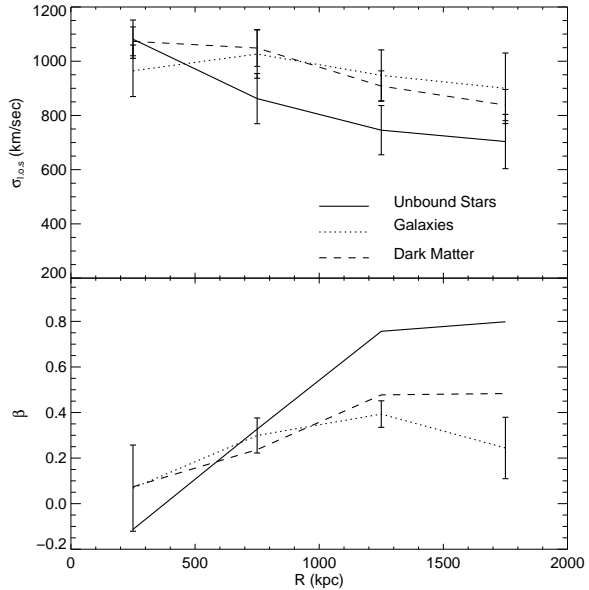}}\\
\end{center}
\caption{Top panel: The line-of-sight velocity dispersion profiles of
the intracluster stars, the dark matter, and the cluster galaxies.
$\beta$ $= -\infty$ corresponds to circular orbits, = 0 corresponds to
isotropic orbits, and = 1 corresponds to radial orbits. Bottom panel:
The 3D anisotropy profile.  Cluster projection does not affect the 3D
$\beta$ profile.  Error bars are not plotted for dark matter and
unbound stars, because the formal bootstrapped errors are tiny, due to
the large number of data points in each bin.}
\label{fig:betaprofiles}
\end{figure}

The top panel of Figure~\ref{fig:betaprofiles} shows the 1D radial
profiles of the line-of-sight (los) velocity dispersions of IC stars,
cluster galaxies, and cluster dark matter particles. To derive the
error bars, we did a bootstrap resampling of the data for numerous
cluster orientations.  We find that the intracluster stars have a
velocity dispersion on average $\sim20\%$ smaller than that of either
the galaxies or the dark matter.  The IC stars' velocity dispersion
decreases with radius a bit more than that of the dark matter or the
galaxies.  This difference in velocity profile is due to the fact
that, beyond 750 kpc, the intracluster star orbits are more radial
than those of either the dark matter or the galaxies. This bias is
seen in the comparison between the anisotropy profiles ($\beta = 1 -
<\sigma^2_t>/<\sigma^2_r>$) of IC stars, galaxies, and dark matter in
the lower panel of Figure~\ref{fig:betaprofiles}.  Although
\citet{sommerlarsen04} did not find such an anisotropy bias in their
simulations, it is not unexpected for unbound stars to have more
radial orbits than galaxies or dark matter; stars are preferentially
stripped from galaxies on radial orbits because, on average, such
orbits have smaller pericenters than isotropic orbits
\citep{taffoni03}.  This bias in the anisotropy of intracluster stars,
relative to that of either cluster galaxies or the dark matter, needs
to be considered when constructing dynamical models based on
intracluster stellar tracers.  In the next section, we will do a
simple calculation to demonstrate the possible effect of this
difference on cluster mass estimates.

\section{Intracluster Stars as a Tracer of Cluster Mass}
The number of intracluster PNe easily surpass that of cluster
galaxies, which suggests that PNe may be a more effective tracer of
the cluster mass distribution than galaxies \citep{ford02}.  Modeling
the mass of a system based on the observed light and line of sight
velocity dispersion profile relies on assumptions about either the
mass distribution or the distribution of orbits.  Therefore, the
extent to which free floating stars can be used as mass tracers is
determined by the extent to which we understand these properties of
the intracluster population relative to that of the underlying mass
distribution. 

In \S4.2, we showed that IC stars are correlated in phase space and
have a mean orbital distribution that differs from that of the cluster
galaxies and dark matter.  Furthermore, samples of galaxy cluster PNe
are identified in discrete fields, such as those of
\citet{arnaboldi02} and \citet{feldmeier98}.  In this section, we use
a crude approach to explore if PNe may nonetheless be useful tracers
of cluster mass.

The ``projected virial theorem'' is one simple dynamical method
typically used to derive galaxy cluster masses \citep{girardi98}.
However, numerous close pairs of objects will be present in samples of
Planetary Nebulae identified in discrete fields, resulting in a
projected virial mass that is more than an order of magnitude less
than the actual cluster mass.  We therefore use the ``projected mass''
to evaluate Planetary Nebulae as mass tracers, because it is less
sensitive to close pairs of objects \citep{rines03}. We computed the
projected mass (PM) with the relation from \citet{heisler85}:

\begin{equation}
M_V = \frac{32}{\pi G} \sum_{i=1}^n \frac{R_i (v_i - v)^2}{N}
\end{equation}

\noindent that assumes isotropic orbits and a continuous mass
distribution \citep{rines03}.  $R_i$ is each object's projected
distance from the cluster center, $v_i$ is each objects's observed
line-of-sight velocity, and $v$ is the average line-of-sight velocity.
The factor of 32 becomes 16 or 64 for circular or radial orbits,
respectively.

\begin{figure}
\begin{center}
\resizebox{8cm}{!}{\includegraphics{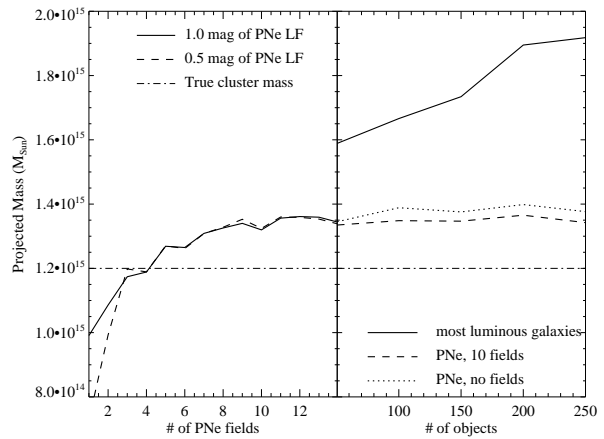}}\\
\end{center}
\caption{Left panel: Projected mass of the galaxy cluster calculated
using a simulated sample of Planetary Nebulae selected in discrete
fields between a projected radius of 500 and 1000 kpc, using either the
brightest 0.5 mag or the brightest 1.0 mag in each of up to 14
fields. Right panel: Projected mass calculated using the N brightest
galaxies, N randomly selected unbound stars, or N simulated Planetary
Nebulae distributed between 14 0.5 x 0.5 degree$^2$ fields.}
\label{fig:cluster_mass}
\end{figure}

First, we calculated the projected mass with PNe from a number of
30$'$ x 30$'$ fields ranging from 1 -- 14, and with a spectroscopic
depth of either the brightest 0.5 or 1.0 magnitudes of the PNLF.  We
randomly placed 40 fields at distances between 0.5 and 1.0 Mpc from
the cluster center and automatically selected those fields farthest
from any galaxy.  The projected mass was thus calculated for numerous
cluster projections.  The average mass as a function of number of
fields is shown in the left panel of
Figure~\ref{fig:cluster_mass}. This plot shows that the number of PNe
fields used in the analysis is more important than the depth of the
spectroscopic follow up, for purposes of calculating mass. The scatter
due to cluster projection is $\sim$25\% of the total cluster mass, similar
to the dispersion found by \citet{governato01}.  Varying the
clustercentric radii of the PNe fields showed that fields at a wide
range of radii need to be included to avoid over/underestimates on the
order of 20\% the mean calculated value. Coma is larger than Virgo, so
it may be that PNe in fields such as \citet{arnaboldi02}'s RCN1 and
fields 2 and 6 of \citet{feldmeier03} are at sufficiently large radii
to be included in an accurate mass measurement.

The projected mass calculated with 5 or more fields nearly asymptotes
to the value calculated using unbound stars not restricted to
individual fields ($1.4 \cdot 10^{15} M_{\odot}$), shown in the right
panel of Figure~\ref{fig:cluster_mass}.  Surprisingly, this calculated
PM is within 10$\%$ of the actual cluster mass of $1.5 \cdot 10^{15}$.
This result is initially surprising because Figure 10 shows both that
IC stars have a velocity dispersion that is somewhat smaller than that
of the underlying mass distribution and that IC stars do not tend to
follow isotropic orbits. These factors produce a mass underestimate
relative to an isotropic population with the same velocity dispersion
as the underlying mass distribution, when using Equation 1.  However,
this underestimate is canceled out by the fact that Equation 1
systematically produces an overestimate of mass relative to true
cluster mass, similar to the overestimate produced by the projected
virial theorem without a surface term correction
\citep{aceves99,rines03}.  We found this same overestimate when we
calculated the average projected mass with C2 cluster galaxies, as
seen in the right panel of Figure~\ref{fig:cluster_mass}. The
1$\sigma$ scatter due to cluster projection is $\sim 15\%$ for the
masses calculated using galaxies. We also found that the projected
masses calculated using galaxies are very similar to the projected
virial masses without a correction for the surface term.

Our analysis shows that IC PNe are worth a closer investigation into
their utility in making detailed mass models of nearby clusters.  The
fact that our analysis yielded a cluster mass very close to the true
mass was partially a fortuitous canceling of two effects; a detailed
analysis would thus likely produce a more reliable approach to cluster
mass modeling with individual IC stars than we have presented here.

\section{Conclusion}
In this paper, we used a high-resolution, SPH simulation of a
Coma-like cluster formed in a cosmological context to study the
formation, evolution and properties of intracluster stars in a rich
galaxy cluster.  This simulation resolves galaxies as faint as $M_R =
-19$, which includes over 85$\%$ of the light from cluster galaxies.
Furthermore, the simulated stellar distributions have a range of effective radii
that matches that observed by Bernardi et al. (2003).  This
dynamic range and accurate stellar distribution are essential for
relying on simulated stars as accurate tracers of the dynamically
stripped stellar population.

Overall, we find that the fraction and distribution of IC stars in a
hierarchical scenario are a good match to the observed properties,
qualitatively similar to the results found in the lower resolution
simulations of \citet{murante04} and \citet{sommerlarsen04}.  $\sim
20\%$ of stars are unbound to any galaxy at z = 0.  The ICL
accumulation is an ubiquitous, ongoing process, with an unbound stars
fraction that slowly increases with time.  The intracluster population
is formed via both stripping within the cluster and via the infall of
large groups that contain stars that are already unbound.  As a
result, there is a link between the accumulation of IC stars and
infall into the cluster. However, the actual IC star fraction at any
given time is not necessarily correlated with the dynamical state of
the cluster, with both the lowest and highest IC fractions occurring
after substantial merging events.

Our analysis traced the fraction of stars stripped
from galaxies as a function of galaxy luminosity.  We found that the
most luminous galaxies in a cluster contribute a substantial fraction
of stars to the unbound population, and that the stripping efficiency
is roughly constant over a wide range of galaxy masses.  As a result,
we predict that IC stars will have photometric properties in common
with the average properties of intermediate luminosity galaxies, which
is consistent with existing observations (e.g. Durrell et al 2002).

We verified the phase space and velocity distribution of IC stars
found in previous numerical work.  IC stars exhibit non-Gaussian
velocity distributions, as well as visible correlations in their
observable phase space plots \citep{napolitano03}.  We also showed
that the orbits of IC stars are more radially biased than those of the
overall galaxy or dark matter, which results in the IC stars having a
lower l.o.s. velocity dispersion in the outskirts of the cluster than
in its inner regions.

We find that despite the fact that IC stars do exhibit phase space
substructure, and have l.o.s. velocities and anisotropies that differ
from that of the underlying distribution, they are promising tracers
of galaxy cluster mass.  The depth of the spectroscopic observations
in each field makes little difference in the calculated masses.  The
projected masses calculated with discrete fields of unbound stars
quickly approaches that calculated with randomly selected unbound
stars, which is within 10$\%$ of the true cluster mass due to a
fortuitous canceling of two effects.  Our analysis suggests that more
detailed analysis may provide a ``correction'' to the projected mass
found with IC stars, similar to the known correction for the projected
virial mass estimate.

Much progress remains to be done to complete our understanding of
intracluster stars and what they can tell us about the formation
history of the clusters which they inhabit.  Some groups are already
(e.g. Mihos et al 2004) starting to include galaxies with more
realistic profiles into numerical simulations, which will enable a
more quantitative analysis of substructure predictions and of
stripping efficiencies. Better resolution coupled with a more realistic
implementation of feedback should be the goal of future cosmological
simulations used to study the properties of intracluster
stars.  These improvements will enable more robust predictions of
colors, gradients and light profiles than possible with the current
generation of simulations. 


\section*{Acknowledgments}
We would like to thank Anthony Gonzalez, Andrey Kravtsov, and Japser
Sommer-Larsen for helpful comments on this manuscript.  BW would like
to thank George Lake for helpful conversations during this project. BW
would like to acknowledge support from the UW Royalty Research Fund,
NSF Grant AST-0098557, and NSF grant AST-0205413.  FG is a Brooks Fellow
and was supported in part by NSF grant AST-0098557.  TRQ acknowledges
support from NSF grant AST-0098557 and NSF grant AST-0205413.  The
simulations were run at the Arctic Region Supercomputing Center.




\end{document}